\documentclass[aps,prc,twocolumn,superscriptaddress,showpacs]{revtex4-1}
\usepackage{graphicx}

\begin{document}


\title{Molecular structures in T=1 states of $^{10}$B}


\author{A.N. Kuchera}
\affiliation{Department of Physics, Florida State University, Tallahassee, Florida 32306-4350}

\author{G.V. Rogachev}
\email{grogache@fsu.edu}
\affiliation{Department of Physics, Florida State University, Tallahassee, Florida 32306-4350}

\author{V.Z. Goldberg}
\affiliation{Cyclotron Institute, Texas A\&M University, College Station, Texas} 

\author{E.D. Johnson}
\affiliation{Department of Physics, Florida State University, Tallahassee, Florida 32306-4350}

\author{S. Cherubini}
\affiliation{Istituto Nazionale di Fisica Nucleare - Laboratori Nazionali del Sud, Catania, Italy}
\affiliation{Dipartimento di Fisica e Astronomia, Universit$\grave{a}$ di Catania, Italy}

\author{M. Gulino}
\affiliation{Istituto Nazionale di Fisica Nucleare - Laboratori Nazionali del Sud, Catania, Italy}
\affiliation{Dipartimento di Fisica e Astronomia, Universit$\grave{a}$ di Catania, Italy}

\author{M. La Cognata}
\affiliation{Istituto Nazionale di Fisica Nucleare - Laboratori Nazionali del Sud, Catania, Italy}

\author{L. Lamia}
\affiliation{Istituto Nazionale di Fisica Nucleare - Laboratori Nazionali del Sud, Catania, Italy}
\affiliation{Dipartimento di Fisica e Astronomia, Universit$\grave{a}$ di Catania, Italy}

\author{S. Romano}
\affiliation{Istituto Nazionale di Fisica Nucleare - Laboratori Nazionali del Sud, Catania, Italy}
\affiliation{Dipartimento di Fisica e Astronomia, Universit$\grave{a}$ di Catania, Italy}

\author{L.E. Miller}
\affiliation{Department of Physics, Florida State University, Tallahassee, Florida 32306-4350}

\author{R.G. Pizzone}
\affiliation{Istituto Nazionale di Fisica Nucleare - Laboratori Nazionali del Sud, Catania, Italy}

\author{G.G. Rapisarda}
\affiliation{Istituto Nazionale di Fisica Nucleare - Laboratori Nazionali del Sud, Catania, Italy}
\affiliation{Dipartimento di Fisica e Astronomia, Universit$\grave{a}$ di Catania, Italy}

\author{M.L. Sergi}
\affiliation{Istituto Nazionale di Fisica Nucleare - Laboratori Nazionali del Sud, Catania, Italy}
\affiliation{Dipartimento di Fisica e Astronomia, Universit$\grave{a}$ di Catania, Italy}

\author{C. Spitaleri}
\affiliation{Istituto Nazionale di Fisica Nucleare - Laboratori Nazionali del Sud, Catania, Italy}
\affiliation{Dipartimento di Fisica e Astronomia, Universit$\grave{a}$ di Catania, Italy}

\author{R.E. Tribble}
\affiliation{Cyclotron Institute, Texas A\&M University, College Station, Texas} 

\author{W.H. Trzaska}
\affiliation{Physics Department, University of Jyv\"askyl\"a, Jyv\"askyl\"a, Finland}

\author{A. Tumino}
\affiliation{Istituto Nazionale di Fisica Nucleare - Laboratori Nazionali del Sud, Catania, Italy}
\affiliation{Dipartimento di Fisica e Astronomia, Universit$\grave{a}$ di Catania, Italy}


\date{\today}

\begin{abstract}
\begin{description}
\item[Background] Multi-center (molecular) structures can play an important role in light nuclei. The highly deformed rotational band in $^{10}$Be with band head at 6.179 MeV  has been observed recently and suggested to have an exotic $\alpha$:2n:$\alpha$ configuration.
\item[Purpose] Search for states with $\alpha$:pn:$\alpha$ two-center molecular configurations in $^{10}$B that are analogous to the states with $\alpha$:2n:$\alpha$ structure in $^{10}$Be.
\item[Method] The T=1 isobaric analog states in $^{10}$B were studied in the energy range of $E_{x}=8.7-12.1$ MeV using the reaction $^1$H($^9$Be,$\alpha$)$^6$Li*(T=1, 0$^+$, 3.56 MeV). 
An R-matrix analysis was used to extract parameters for the states observed in the (p,$\alpha$) excitation function.
\item[Results] Five T=1 states in $^{10}$B have been identified. The known 2$^+$ and 3$^-$ states at 8.9 MeV have been observed and their partial widths have been measured. The spin-parities and partial widths for three higher lying states were determined.
\item[Conclusions] Our data support theoretical predictions that the 2$^{+}$ state at 8.9 MeV (isobaric analog of the 7.54 MeV state in $^{10}$Be) is a highly clustered state and can be identified as a member of the $\alpha$:np:$\alpha$ rotational band. The next member of this band, the 4$^+$ state, has not been found. A very broad 0$^+$ state at 11 MeV that corresponds to pure $\alpha$+$^6$Li(0$^+$,T=1) configuration is suggested and it might be related to similar structures found in $^{12}$C, $^{18}$O and $^{20}$Ne.
\end{description}
\end{abstract}

\pacs{21.10.-k, 27.20.+n, 25.40.Ny}

\maketitle

\section{Introduction}
Clustering phenomena clearly manifest themselves in light nuclei. It is well established that the low-lying states in $^8$Be can be described as a two-center $\alpha$-$\alpha$ structure \cite{Hiura1972}. Recent {\it ab initio} Green's Function Monte Carlo calculations explicitly show how this structure emerges naturally for the $^8$Be ground state \cite{Wiringa2000}. The suggestion that this two-center structure may survive when ``valence'' nucleons are added to the system has been made in the early 1970s \cite{Hiura1972}. A semi-quantitative discussion of this subject can be found in \cite{Oertzen1996} where the two-center molecular states in $^9$B, $^9$Be, $^{10}$Be, and $^{10}$B nuclei were considered in the framework of a two-center shell model. An Antisymmetrized Molecular Dynamics plus Hartree-Fock (AMD+HF) approach was proposed in \cite{Dote1997} as a theoretical tool to study the structure of low-lying levels in $^{9,10,11}$Be isotopes. Deformation (distance between the two $\alpha$s) for several low-lying states in Be isotopes has been studied. Very large deformation ($\beta$=0.852) for the 6.179 MeV 0$^+$ state in $^{10}$Be was suggested, which corresponds to an interalpha distance of 3.55 fm. This is 1.8 times more than the corresponding value for the $^{10}$Be ground state. A similar result was obtained in \cite{Itagaki2000} where the spectrum of $^{10}$Be was reasonably well reproduced using a molecular orbit (MO) model. The second 0$^+$ state in $^{10}$Be has an enlarged $\alpha$-$\alpha$ distance and the highly deformed rotational band with large moment of inertia built on that configuration emerges according to these calculations. The two known states in $^{10}$Be, the 0$^+$ at 6.179 MeV and the 2$^+$ at 7.542 MeV are believed to be associated with this rotational band. 

The experimental evidence for the 4$^+$ member of this band is controversial. A state at excitation energy of 10.2 MeV was first observed in the $^7$Li($\alpha$,p) reaction \cite{Hamada1994} and then in $^7$Li($^7$Li,$\alpha$) where it was found that the main decay channel for this state is $\alpha$+$^6$He \cite{Soic1996}. The spin-parity of this state was first determined in \cite{Curtis2001} using angular correlations between the $^7$Li($^7$Li,$\alpha$$^6$He)$\alpha$ reaction products and was found to be 3$^-$ (excitation energy of this state was revised to 10.15 MeV). This would clearly disqualify the 10.15 MeV state as a member of the highly deformed $K^{\pi}$=0$^+$ rotational band. However, two more recent results put this spin-parity assignment into question. The $^6$Li($^6$He,$\alpha$$^6$He)$^2$H reaction was used to populate excited states in $^{10}$Be where the 10.15 MeV state was observed and angular correlations between the reaction products seemed to indicate a 4$^+$ spin-parity assignment \cite{Milin2005}. The state was also studied in $\alpha$+$^6$He resonance scattering and the 4$^+$ spin-parity assignment was confirmed \cite{Freer2006}. The partial $\alpha$ width of the state determined in \cite{Freer2006} was found to be very large, approaching or exceeding the $\alpha$ single-particle limit. It was argued in \cite{Freer2006} that the 10.15 MeV state is the next member of the highly deformed $K^{\pi}$=0$^+$ rotational band built on the 0$^+$ state at 6.179 MeV and that the very large moment of inertia of this band indicates $\alpha$:2n:$\alpha$ configuration. The 6$^+$ member is also predicted for this band in \cite{Wolsky2010}.
 
An analogous band with $\alpha$:np:$\alpha$ configuration should be found in the spectrum of T=1 states in $^{10}$B. Isobaric analogs of the first two members of this band in $^{10}$B are known, these are the T=1 0$^+$ at 7.56 MeV and T=1 2$^+$ at 8.89 MeV. The goal of this work was to study the quasi-molecular T=1 states in $^{10}$B with $\alpha$+$^6$Li(T=1) structure using the $^1$H($^9$Be,$\alpha$)$^6$Li*(T=1, 0$^+$, 3.56 MeV) reaction.  While both T=0 and T=1 states can be populated in $^9$Be+p scattering, only the T=1 states would predominantly decay to the T=1 state in $^6$Li by $\alpha$ emission due to isospin conservation and the T=0 states would be strongly suppressed in the $^1$H($^9$Be,$\alpha$)$^6$Li*(T=1, 0$^+$, 3.56 MeV) excitation function. One of the main issues is a search for the 4$^+$ member of the hypothetical $\alpha$:np:$\alpha$ rotational band in $^{10}$B. The present study is also a part of the search for finding the best way to populate the $0^+$(T=1) 3.56 MeV level in $^6$Li for subsequent investigation of parity violating $\alpha$-decay\cite{Goldberg2006}.

\section{Experimental details}

The experiment was performed at the Laboratori Nazionali del Sud INFN in Catania, Italy.  A $^9$Be beam was accelerated by the SMP Tandem onto a solid polyethylene (CH$_2$) target. Two different targets were used with thicknesses of 56 and 74 $\mu$g/cm$^2$. Excitation functions of $^1$H($^9$Be,$\alpha$)$^6$Li*(T=1, 0$^+$, 3.56 MeV) were measured by changing the beam energy from 22.3 MeV up to 55.6 MeV with steps ranging between 0.12 and 5.75 MeV (0.012 MeV and 0.575 MeV in c.m.) with typical beam currents $\approx1$ nA. Inverse kinematics (heavy projectile and light target) was chosen to measure the recoils because the higher forward momentum gives a boost in the lab frame needed to measure the otherwise low energy $\alpha$ and $^6$Li decay products.  The 3.56 MeV state in $^6$Li is unbound, but it decays predominantly by $\gamma$ emission. Parity violating $\alpha$+d decay and $\gamma$ decay to the continuum with subsequent $\alpha$+d have branching ratios of $\le6.5\times10^{-7}$ \cite{Robertson1984} and $\sim8\times10^{-5}$ \cite{Grigorenko1998} respectively and can be neglected for the purpose of this work.

\begin{figure}
 \includegraphics[width=3in]{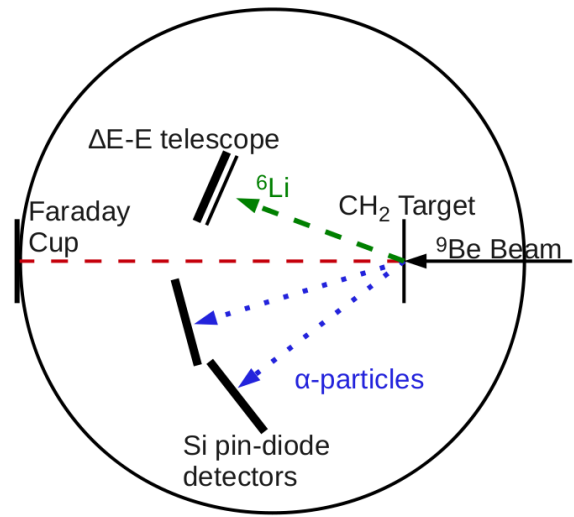}
 \caption{\label{fig:chamber} (Color online) Experimental setup to measure the excitation function for $^1$H($^9$Be,$\alpha$)$^6$Li reaction. The position sensitive $\Delta E-E$ telescope (with gas ionization chamber as $\Delta E$) was used to identify the $^6$Li recoils. $\alpha$ particles were measured in an array of Si pin-diode detectors. A Faraday cup was installed at the back of the chamber to measure the beam current.}
 \end{figure}

The experimental setup is shown in Figure \ref{fig:chamber}. Three arrays of detectors were used. Two arrays of three square silicon pin-diode detectors with active area of $18\times18$ mm$^2$ and thickness of 500 $\mu$m covered lab angles 4.2$^\circ - 19.3^\circ$. These were used as $\alpha$-particle detectors.  A $\Delta E-E$ telescope covered the lab angles 4.75$^\circ - 11.8^\circ$. The telescope consisted of a gas ionization chamber filled with P-10 gas and a position sensitive silicon detector (resistive layer). The active area of the telescope was determined by a collimator 45.9 mm long and 2 mm wide. The $\Delta E-E$ telescope was used to identify and measure the position and energies of the $^6$Li recoils.

 \begin{figure}
 \includegraphics[width=3.2in]{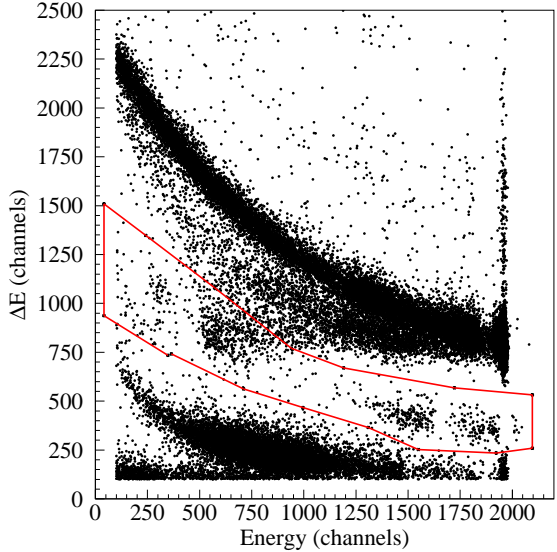}%
 \caption{\label{fig:deltaEE} (Color online) Typical $\Delta E-E$ spectrum measured in the telescope with an example of the cut that was used to select the $^6$Li recoils (red contour). The band above the $^6$Li corresponds to the $^9$Be and below is the $\alpha$ band.}
 \end{figure}
 
 \begin{figure}
 \includegraphics[width=3.2in]{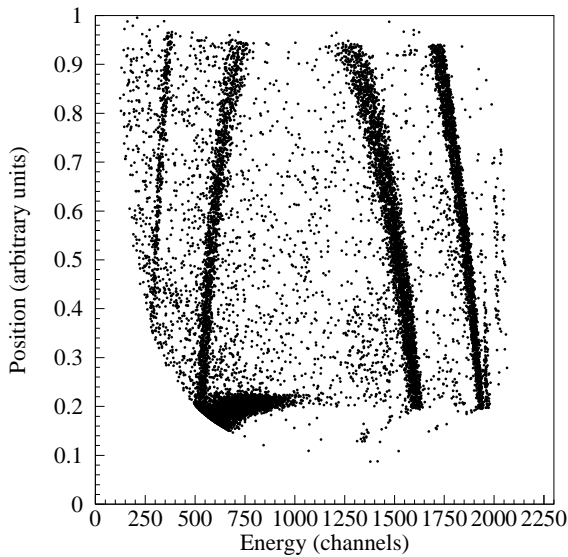}%
 \caption{\label{fig:posE_high} Angle-energy correlation plot for $^6$Li recoils measured at E$_{cm}= 4.68$ MeV by position sensitive $\Delta$E-E telescope. The inner arch corresponds to the $^1$H($^9$Be,$\alpha$) reaction that leaves $^6$Li in the 3.56 MeV T=1 $0^+$ excited state. The outer arch corresponds to the population of the $^6$Li ground state. The two branches for each arch are due to forward and backward scattering in the c.m.}
 \end{figure}

The $^1$H($^9$Be,$\alpha$)$^6$Li*(T=1, 0$^+$, 3.56 MeV) events were selected using kinematic coincidences in two sets of detectors or by using angle-energy correlation of $^6$Li ions preselected using $\Delta E-E$ 2-D spectra as shown in Figure \ref{fig:deltaEE}. Due to inverse kinematics, the 2-D angle-energy correlation plot shows an outer and inner arc (Fig. \ref{fig:posE_high}). The outer arc corresponds to the events in which the $^6$Li was formed in its ground state and the inner arc corresponds to the $^6$Li populated in its $0^+$ 3.56 MeV state. The arcs are formed due to inverse kinematics for which $^6$Li scattered backward in the center-of-momentum frame still goes forward in the lab frame since c.m. has large total forward momentum.

\begin{figure}
 \includegraphics[width=3.45in]{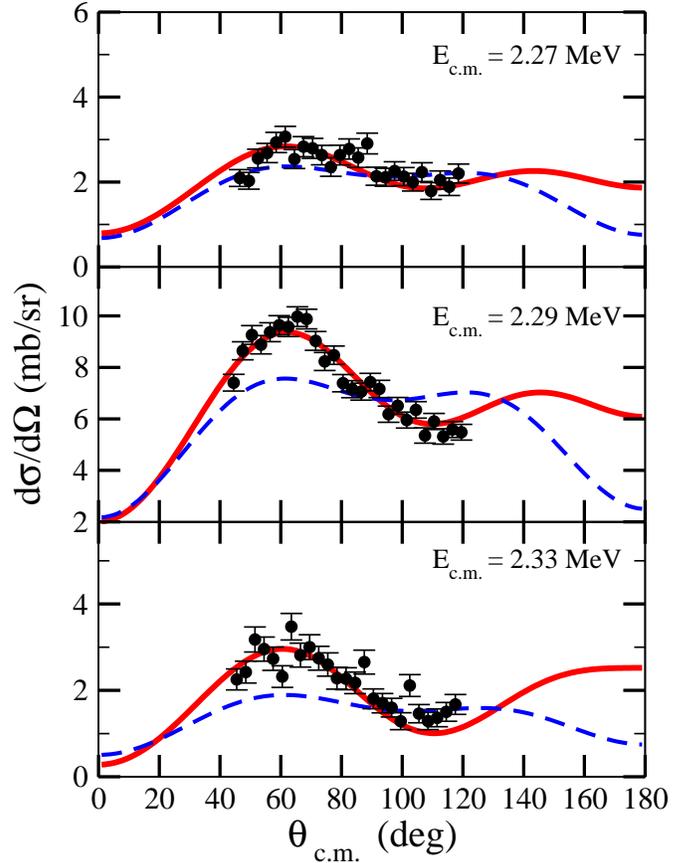}%
 \caption{\label{fig:angdislow} (Color online) Angular distributions of the 2$^+$/3$^-$ doublet at E$_{cm}=2.3$ MeV (E$_x$=8.9 MeV). The three energies are 20 keV below the maximum cross section (top), near the maximum of the cross section (center), and 14 keV above the maximum of the cross section (bottom). Solid (red) line is the best R-matrix fit and dashed (blue)  line is the R-matrix calculation with small (25 eV) partial alpha width in the 3$^-$ state.}
\end{figure}

Measurement of absolute cross section was handicapped by significant target deterioration observed during the run. Therefore, the absolute normalization of the cross section was performed by evaluating the product of the effective target thickness and integral of accumulated $^9$Be beam ions
from the simultaneously measured $^1$H($^9$Be,p)$^9$Be elastic scattering cross section known from previous experiments \cite{Yasue1972, Allab1983}. The uncertainty of this procedure was evaluated using 3 pairs of identical beam energy data points that were measured in six separate runs at 22.9 MeV, 37.1 MeV and 38.7 MeV of $^9$Be. It was found that the normalization procedure yields an uncertainty of $\sim$10\%. The $^9$Be(p,p)$^9$Be cross section was measured with systematic uncertainty of 13\% in \cite{Allab1983} and this is carried over as systematic uncertainty of the absolute cross section determined here. Statistical uncertainty for the excitation functions shown in Figs \ref{fig:55deg},\ref{fig:zeroplus},\ref{fig:exfuncompare} are $\sim$2 \% (binning of 5$^{\circ}$ in c.m. was used) and the error bars are dominated by the uncertainty in the normalization procedure (systematic uncertainty is not included in the error bars). Uncertainties shown in angular distribution figures (Figs \ref{fig:angdislow},\ref{fig:zeroplusangdis},\ref{fig:angdishi}) are statistical uncertainties of these measurements.

\section{Results}
\begin{figure}
 \includegraphics[width=3.45in]{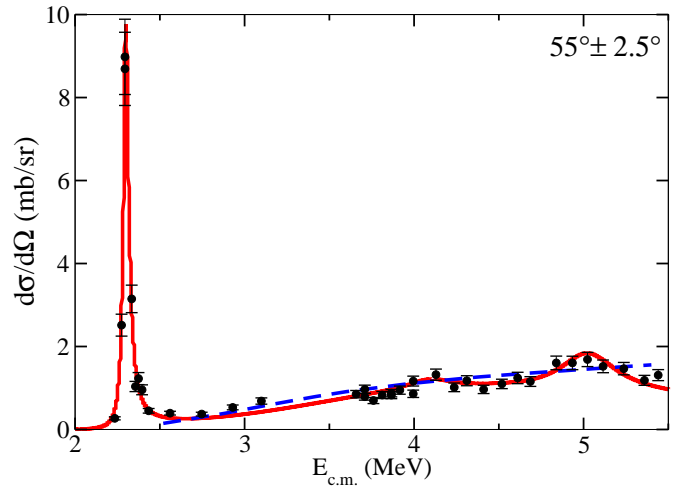}%
 \caption{\label{fig:55deg} (Color online) $^1$H($^9$Be,$\alpha$)$^6$Li*(T=1, 0$^+$, 3.56 MeV) excitation function at $55^{\circ}$ in c.m. Solid (red) curve is the best R-Matrix fit. The R-matrix parameters for this fit are listed in Table \ref{tab:all}. The dashed (blue) curve is the cross section due to direct (p,$\alpha$) process calculated using the DWBA code FRESCO assuming unity for the overlap integral (spectroscopic factor) between $^9$Be(g.s.) and the $^6$Li$(0^+)\times^3$H configuration.}
 \end{figure}

The excitation functions for the $^1$H($^9$Be,$\alpha$)$^6$Li*(T=1, 0$^+$, 3.56 MeV) reaction were measured in the c.m. energy range between 2.1 and 5.5 MeV (8.7 - 12.1 MeV excitation in $^{10}$B). The excitation function at $55\pm2.5^{\circ}$ in c.m. is shown in Figure \ref{fig:55deg}. The narrow peak at E$_{cm}=2.3$ MeV corresponds to the doublet of known T=1 states in $^{10}$B, the 3$^-$ and the 2$^+$ at E$_x=8.9$ MeV. While spin-parities and widths of these states are known, the decay branching ratios are not. The data from this work allow us to determine the partial widths for these states. This is particularly important for the 2$^+$ state at E$_x=8.89$ MeV to verify whether it is a member of the highly deformed $\alpha$:np:$\alpha$ rotational band.

A multi-channel, multi-level R-matrix analysis of the excitation functions was performed. The three dominant decay channels included in the R-matrix fit are the p+$^9$Be(g.s.), n+$^9$B(g.s.) and $\alpha$+$^6$Li(0$^+$, 3.56 MeV) channels. $\alpha$ decay to the T=0 states in $^6$Li does not conserve isospin and it was assumed to be negligible. Though $\alpha$ decay to $^6$Li(g.s.) was included for the $2^+$ state at E$_{cm}=2.3$ MeV as there is clear evidence for small (less than 4 keV) $\Gamma_{\alpha_0}$ partial widths in this resonance \cite{Weber1954}. Another isospin violating channel is deuteron decay to the ground state of $^8$Be. This channel has favorable penetrability factors. Therefore, even small violation of isospin symmetry at the level of less than one percent can contribute few keV to the total width of the 2$^+$ state. Contribution from these two channels to the total width of the 2$^+$ state was considered as one free parameter in the R-matrix fit. One more open channel is proton decay to the first excited state in $^9$Be (the 1/2$^+$ at 1.7 MeV), but it cannot compete with the decay to the $^9$Be g.s. at excitation energies of interest due to small penetrability factors. The same is true for the decay into $^8$Be and the p-n singlet state. The reduced width amplitudes were constrained in the R-matrix fit by requiring that corresponding proton and neutron reduced width amplitudes are equal, as determined by the isospin Clebsch-Gordon coefficients for the T=1 states. The following results are organized in subsections for each resonance (the $2^+$ and $3^-$ states at E$_{cm}=2.3$ MeV are grouped together because of their proximity to each other and because of the crucial role played by interference between these two states).

\subsection{$2^+/3^-$ doublet at $E_{cm}=2.3$ MeV}

The total width of the 2$^+$ state at 8.894 MeV in $^{10}$B produced by our best fit is 34$\pm$4 keV (Table \ref{tab:all}). This is 6 keV lower than the 40$\pm$1 keV given in \cite{Kiss1977} but it agrees very well with earlier $^9$Be(p,n) data \cite{Marion1956} where the width of this state was determined to be 34$\pm$3 keV. The angular distributions near E$_{cm}=2.3$ MeV (E$_x=8.9$ MeV in $^{10}$B) are shown in Figure \ref{fig:angdislow}. It was found that the R-matrix fit of the excitation functions near E$_{cm}$=2.3 MeV is not unique. Two different solutions related to the parameters of the 2$^+$ state were identified. One set of parameters has the partial proton width larger than the partial $\alpha$ width (22 keV and 6 keV respectively), and the other set has larger partial $\alpha$ width ($\Gamma_\alpha$=18 keV) than the proton width ($\Gamma_p$=7 keV). Both options fit the experimental data equally well (cross section is proportional to the product of these partial widths) and it is not possible to select between them based on R-matrix analysis alone. However,  when the widths of the states are converted from the $^{10}$B to the $^{10}$Be system using isospin invariance, the larger proton width solution produces 23 keV total width for the 2$^+$ state at 7.542 MeV in $^{10}$Be. This is a factor of 3.6 larger than the known experimental value of 6.3$\pm$0.8 keV \cite{Tilley2004}. The smaller proton width solution gives 8$\pm$2 keV in good agreement with the $^{10}$Be data. Therefore, we conclude that partial $\alpha$ width accounts for about 50\% of the natural width of the 2$^+$ state at 8.9 MeV in $^{10}$B (see Table \ref{tab:all}) and has value of 18$\pm$2.0(stat\&norm)$\pm$2.3(sys) keV. It corresponds to 100\% of the $\alpha$ single particle limit, determined by $\theta_{\alpha}^2=\gamma^2_{\alpha}/\gamma^2_{sp}$ ($\gamma^2_{sp}=\hbar^2/(\mu R^2)$, where $\mu$ is the reduced mass and R is the channel radius of 4.77 fm). This provides clear evidence that this state has an extreme $\alpha$-cluster structure. This state also features 12$\pm$5 keV width due to isospin violating $\alpha+^6$Li(g.s.) and d+$^8$Be decay channels. The inclusion of this width into the R-matrix fit is necessary to reproduce the known total width of the state and the measured cross section simultaneously.

It is instructive to compare our results for the 2$^+$/3$^-$ doublet with the results from \cite{Kiss1977}, where the $^9$Be(p,$\alpha$)$^6$Li$^*$(0$^+$, T=1, 3.56 MeV) reaction was studied with high resolution in narrow intervals of c.m. energies from 2.25 to 2.38 MeV. Angular distributions of $\alpha$ particles from this reaction were extracted from the shape of the Doppler shifted $\gamma$-rays from the decay of the $^6$Li$^*$(0$^+$, T=1, 3.56 MeV) state at 17 different c.m. energies. While it was not possible to determine partial $\alpha$ widths since the absolute magnitude of the cross section was not measured, the analysis of the angular distributions constrained some properties of the 2$^+$/3$^-$ doublet. Specifically, it was determined that the ratio $(\Gamma_p{\Gamma}_{\alpha})^{2^+}/(\Gamma_p{\Gamma}_{\alpha})^{3^-}=109\pm20$. Partial widths extracted from our fit yield $(\Gamma_p{\Gamma}_{\alpha})^{2^+} /(\Gamma_p{\Gamma}_{\alpha})^{3^-}=3.0\pm$0.2. This ratio, which is much smaller than in \cite{Kiss1977}, is forced by the necessity to reproduce the angular distribution asymmetry with respect to 90$^{\circ}$ that is evident in the experimental angular distribution in Fig. \ref{fig:angdislow}. The magnitude of the cross section at 2.3 MeV is determined by the product of the partial $\alpha$ and proton widths of the 2$^+$ state, and the nucleon widths are dominant in the $3^-$ and are fixed by the total width of the state. Therefore, the partial $\alpha$ width of the 3$^-$ state is the only free parameter available to modify the shape of the angular distribution. It has to be as small as 22 eV to be consistent with the $(\Gamma_p{\Gamma}_{\alpha})^{2^+}/(\Gamma_p{\Gamma}_{\alpha})^{3^-}$ ratio given in \cite{Kiss1977} but this would produce an almost symmetric angular distribution at 2.3 MeV shown as a dashed (blue) curve in Fig. \ref{fig:angdislow}. The best fit value for the partial $\alpha$ width of the 3$^-$ state is 570$\pm$50 eV. This value corresponds to a rather high degree of clustering for the 3$^-$ state ($\theta^2_{\alpha}=0.42\pm0.04$ for a channel radius of 4.77 fm).

\subsection{\label{sec:0+} Flat region between $E_{cm}=3.7-4.5$, direct reaction or new resonance?}

\begin{figure}
 \includegraphics[width=3.45in]{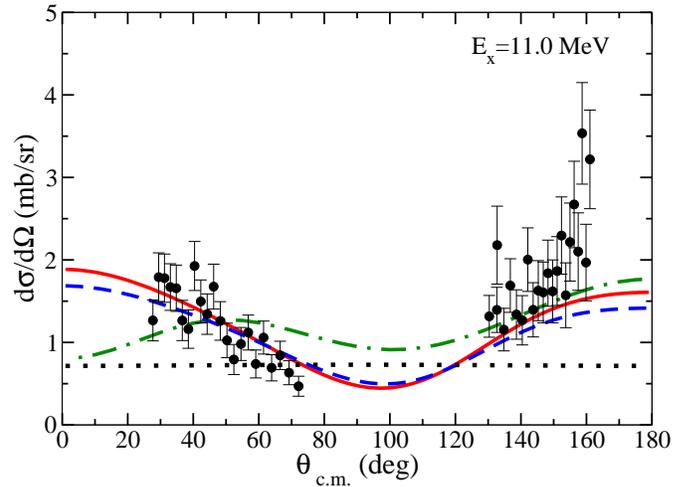}%
 \caption{\label{fig:zeroplusangdis} (Color online) The angular distribution at E$_{cm}=4.4$ MeV (E$_x=11$ MeV). The contribution of the $0^+$ state alone is shown by the dotted (black) line. The angular distribution produced by DWBA calculations at this energy is shown by the dash-dotted (green) line. The dashed (blue) line is the best R-matrix fit with parameters given in the Table \ref{tab:all}. The solid (red) curve is the best R-matrix fit with an additional $2^+$ resonance at higher excitation energy (13 MeV) which was found to improve the overall fit.}
 \end{figure}

The excitation function for the $^1$H($^9$Be,$\alpha$)$^6$Li*(T=1, 0$^+$, 3.56 MeV) reaction does not have any narrow structures above E$_{cm}=2.4$ MeV (Fig. \ref{fig:55deg}). The cross section increases gradually between 2.5 MeV and 3.5 MeV and then stays relatively flat up to 4.5 MeV in c.m. with nearly isotropic angular distribution. This shape of the excitation function is characteristic for the direct (p,$\alpha$) reaction. The dashed (blue) curve in Fig. \ref{fig:55deg} shows the direct reaction cross section calculated using the DWBA code FRESCO \cite{Thompson88}. The CH89 global nucleon optical model potential was used for the p+$^9$Be channel wavefunction \cite{Varner1991}, and the global optical potential for $\alpha$ particles suggested in \cite{Avrigeanu1994} was applied for the $\alpha$+$^6$Li(0$^+$) partition. Radius and diffuseness for the form-factor Woods-Saxon potentials are r$_{\circ}$=1.2 fm and a=0.7 fm for the p$\times^3$H form-factor and r$_{\circ}$=0.88 fm and a=0.65 fm for $^3$H$\times^6$Li$(0^+)$. The smooth energy dependence of the cross section between 2.5 and 4.5 MeV is reproduced by the DWBA calculations. The angular distribution is not too dissimilar from the experimental data (dot-dashed (green) curve in Fig. \ref{fig:zeroplusangdis}). Therefore, energy and angular dependence of the cross section is generally consistent with the direct (p,$\alpha$) process. However, the absolute magnitude of the cross section requires a large value (unity) for the overlap integral (spectroscopic factor) between $^9$Be(g.s.) and the $^6$Li$(0^+)\times^3$H configuration. This is unrealistic and one should expect to see much smaller cross section due to the direct (p,$\alpha$) reaction than is observed experimentally. High cross section is indicative of the presence of resonances. Nevertheless, due to uncertainties in the DWBA model parameters we cannot rule out a direct reaction mechanism as a viable explanation for the measured cross section between 2.5 and 4.5 MeV.

\begin{figure}
 \includegraphics[width=3in]{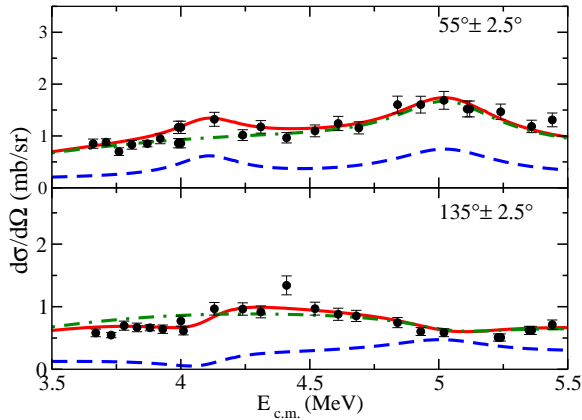}%
 \caption{\label{fig:zeroplus} (Color online) Effect of the broad $0^+$ state at 11 MeV and narrow $2^+$ resonance at 10.7 MeV. The solid (red) line is the best R-matrix fit that includes the $0^+$ and $2^+$ states, and the dashed (blue) line is the same fit but excluding the $0^+$ state, and the dot-dash (green) line is the best fit with the $2^+$ state removed.}
 \end{figure}

Another possibility is to describe the shape of the cross section between 2.5 and 4.5 MeV in the framework of an R-matrix approach using a broad low spin T=1 state. A $0^+$ resonance fits the data at all angles. The effect of the $0^+$ is shown in Fig. \ref{fig:zeroplus}. The solid (red) curve is the best fit that includes a highly clustered broad 0$^+$ state at 4.4 MeV (11 MeV excitation energy in $^{10}$B). The dashed (blue) line shows the fit without including the $0^+$ state. The angular distribution at the resonance energy is shown in Fig. \ref{fig:zeroplusangdis}. Interference between the broad 0$^+$ state and the other states included in the R-matrix fit (especially the broad 1$^-$ at 5.04 MeV) reproduce the measured angular distribution (solid (red) curve in Fig. \ref{fig:zeroplusangdis}). The 0$^+$ state alone would produce an isotropic angular distribution (dotted (black) line in Fig. \ref{fig:zeroplusangdis}). The broad 0$^+$ state has not been suggested in $^{10}$B before and has no known analogs in the spectrum of $^{10}$Be. However, total cross section for the analogous $^9$Be(n,$\alpha$)$^6$He reaction shows a very broad asymmetric peak in the corresponding energy interval \cite{Stelson1957,Bass1961} that may be interpreted as manifestation of the broad low spin state in $^{10}$Be. While the broad 0$^+$ state describes the excitation functions at all angles, it is not possible to reliably exclude other interpretations of the observed enhancement in the (p,$\alpha$) cross section (such as contribution from the direct reaction and/or other broad low spin states). Therefore, we consider the 0$^+$ state tentative.

\subsection{Resonance at $E_{cm}=4.1$ MeV}

Addition of a weak resonance at E$_{cm}=4.1$ MeV with width of $\approx300$ keV improves the fit (see Fig. \ref{fig:zeroplus}). This resonance corresponds to the known 10.84 MeV state \cite{Tilley2004} with previously uncertain spin-parity. Our data is consistent with the 2$^+$ spin-parity assignment for this state and its total width is dominated by nucleon partial widths (see Table \ref{tab:all}). Determining resonance parameters with high precision for this weak state was not possible due to low $\chi^2$ sensitivity. Only a lower limit of $\Gamma>180$ keV  could be determined which is in agreement with the $\Gamma=300\pm100$ reported in \cite{Tilley2004}.  We believe that this is the isobaric analog of the known $2^+$ resonance at 9.56 MeV in $^{10}$Be. Based on the reduced widths amplitudes from our R-matrix fit and taking into account isospin Clebsch-Gordon coefficients we estimated the total width of this 2$^+$ state in $^{10}$Be to be $\approx330$ keV. This value seems to be more consistent with the $^9$Be(d,p) data \cite{Anderson1974} that reports a width of 291$\pm20$ keV for this state rather than with the 141$\pm$10 keV value from the $^7$Li($^7$Li,$\alpha$) data \cite{Liendo2002}.

 \begin{figure}
 \includegraphics[width=3.45in]{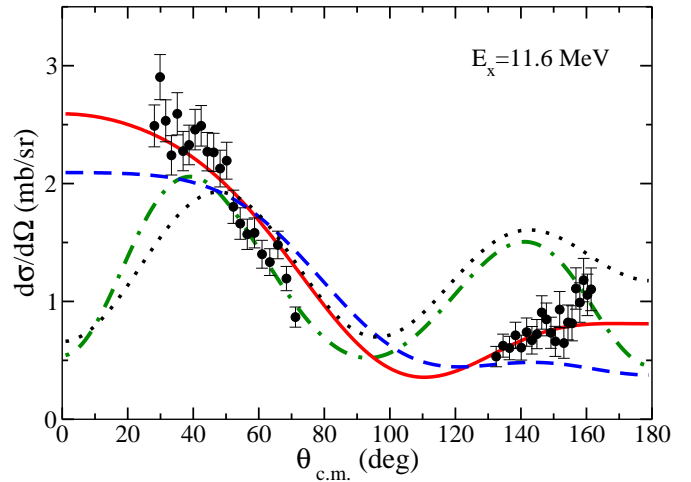}%
 \caption{\label{fig:angdishi} (Color online) Angular distribution at E$_{cm}=5.0$ with four different R-matrix fits. Three of the fits have a different state at 5 MeV. The $4^+$ (dotted, black) and $3^-$ (dash-dot, green) spin-parity assignments show obvious disagreement with the data at large angles. The solid (red) and dashed (blue) fits have a $1^-$ resonance at 5 MeV. The solid (red) curve describes the data best and includes a higher lying $2^+$ state that lifts the cross section at the larger angles.}
 \end{figure}
 
 \subsection{Resonance at $E_{cm}=5.0$ MeV}
 
\begin{figure*}
 \includegraphics[width=6in]{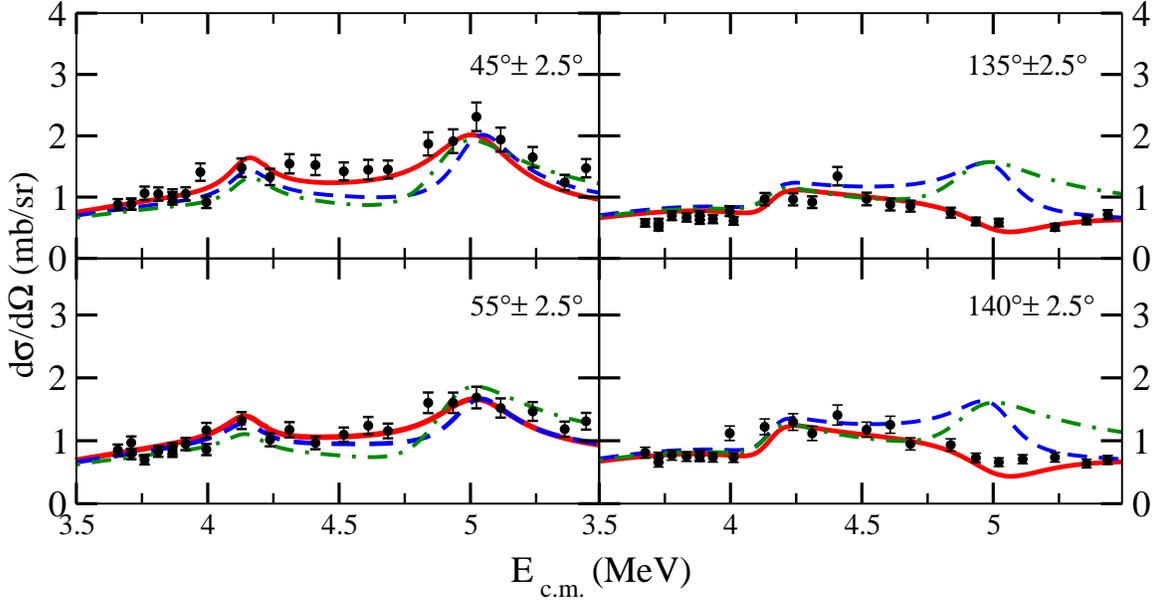}%
 \caption{\label{fig:exfuncompare} (Color online) Excitation functions with three different R-matrix fits that correspond to three spin-parity assignments for the state at 5 MeV. The best fit corresponds to $1^-$ spin-parity assignment, as shown by the solid (red) line, the fit with a $4^+$ spin-parity assignment is shown by the dashed (blue) line, and the 3$^-$ fit is shown by the dot-dash (green) line. All three fits agree with the data relatively well at lower angles, but only the fit with a $1^-$ spin-parity assignment agrees well with larger angle data.}
 \end{figure*}

A structure with width of $\sim$500 keV is evident at E$_{cm}=5.04$ MeV (E$_{x}=11.63$ MeV) (Fig. \ref{fig:exfuncompare}). The shape of the angular distribution shows resonance behavior. It is sharply peaked toward lower c.m. angles and is strongly asymmetric with respect to 90$^{\circ}$ in c.m. indicating interference of states with opposite spin-parity. This structure is intriguing because its excitation energy is very close to the excitation energy of the expected isobaric analog of the 10.15 MeV state in $^{10}$Be. Since the spin-parity assignment for the 10.15 MeV state in $^{10}$Be is 4$^+$ or 3$^-$ we  explored these possibilities systematically. Unfortunately, angular distribution for a specific spin-parity assignment is not unique (except for the 0$^+$ state) due to the fact that neither target nor projectile are spin zero particles and distribution of the populated m-substates in the compound nucleus, $^{10}$B, is not known. We have tried all possible distributions of m-substates as determined by the ratios between the channel spin S=1 and S=2 reduced width amplitudes. The best fits that we were able to achieve for the 4$^+$ and 3$^-$ assignments are shown in Figs. \ref{fig:angdishi} and \ref{fig:exfuncompare} (dotted (black) curve for the 4$^+$ and dot-dashed (green) for the 3$^-$ assignments in Fig. \ref{fig:angdishi}). While it is possible to find a combination of the reduced width amplitudes that fit the low c.m. angles (below 90$^\circ$) reasonably well, the angular distribution at higher c.m. angles cannot be simultaneously reproduced. Therefore, we conclude that neither 4$^+$ nor 3$^-$ spin-parity assignments are consistent with the observed angular distribution. It was found that the fit can be achieved at all angles if a 1$^-$ spin-parity assignment is applied (see Figs. \ref{fig:angdishi} and \ref{fig:exfuncompare}). The total width of this state is dominated by nucleon partial width with dimensionless reduced $\alpha$ width of only 0.005 and is shown by the dashed (blue) curve. The angular distribution fit is improved by the inclusion of a broad $2^+$ state at about 13 MeV excitation energy (1 MeV above the highest excitation energy measured in this experiment). The fit that includes this higher lying $2^+$ state is shown with the solid (red) curve.

\begin{table*}
\caption{\label{tab:all} Resonance parameters determined from the R-matrix fit. $\theta^2_{\alpha}$ was calculated using channel radius of 4.77 fm while $\theta^2_p$ and $\theta^2_n$ correspond to a channel radius of 4.30 fm. Two uncertainties are provided for parameters that are most sensitive to the absolute normalization and for which 13\% systematic uncertainty contributes significantly. The first value corresponds to the combined statistical and relative normalization uncertainty and the second corresponds to systematic uncertainty. The resonance given in parenthesis is tentative.}
\begin{ruledtabular}
\begin{tabular}{cccccccccc}
J$^{\pi}$ & E$_{cm}$ (MeV)& E$_x$ (MeV) & $\Gamma_{tot}$ (keV) & $\Gamma_{\alpha}$ (keV) & $\Gamma_{T=0}$ (keV) & $\Gamma_p$ (keV) & $\Gamma_n$ (keV) & $\theta^2_{\alpha}$ & $\theta^2_p=\theta^2_n$\\
\hline
$2^+$ & 2.308(2) & 8.894(2) & 34(4) & 18$\pm$2.0$\pm$2.3 & 7(4) & 7$\pm$1$\pm$1 & 2(1)& 1.1(2) & 0.031(5)\\
$3^-$  & 2.312(10) & 8.898(10) & 80(10) & 0.57(5) & -& 75(10) & 4(3) &0.42(4) & 0.11(2)\\
$2^+$ & 4.1(1) & 10.7(1) & 300$\pm$100\footnotemark[1] & $\approx$8 &- & $\approx$170  & $\approx$130 & $\approx$0.007 & $\approx$ 0.03   \\
($0^+$) & 4.4(3) & 11.0(3) & 3700$^{+200}_{-600}$ & 2800$^{+200}_{-600}$ &- & 514$\pm$51$\pm$67 & 414$\pm41\pm54$ & 0.97$^{+0.06}_{-0.20}$ & 0.09(2) \\
$1^-$ & 5.04(7) & 11.63(7) & 480(150) & 13(6) &-& 260(100) & 210(120) & 0.004(2) & 0.05(2) \\
\end{tabular}
\end{ruledtabular}
\footnotetext[1]{This value was taken from \cite{Tilley2004}. Sensitivity of $\chi^2$ function with respect to parameters of this weak state is small. We can set a lower limit of $\approx$180 keV but no upper bound could be determined. The total width is dominated by the nucleon widths.}
\end{table*}

\section{Discussion}
\begin{figure*}[htb]
 \includegraphics[width=6in]{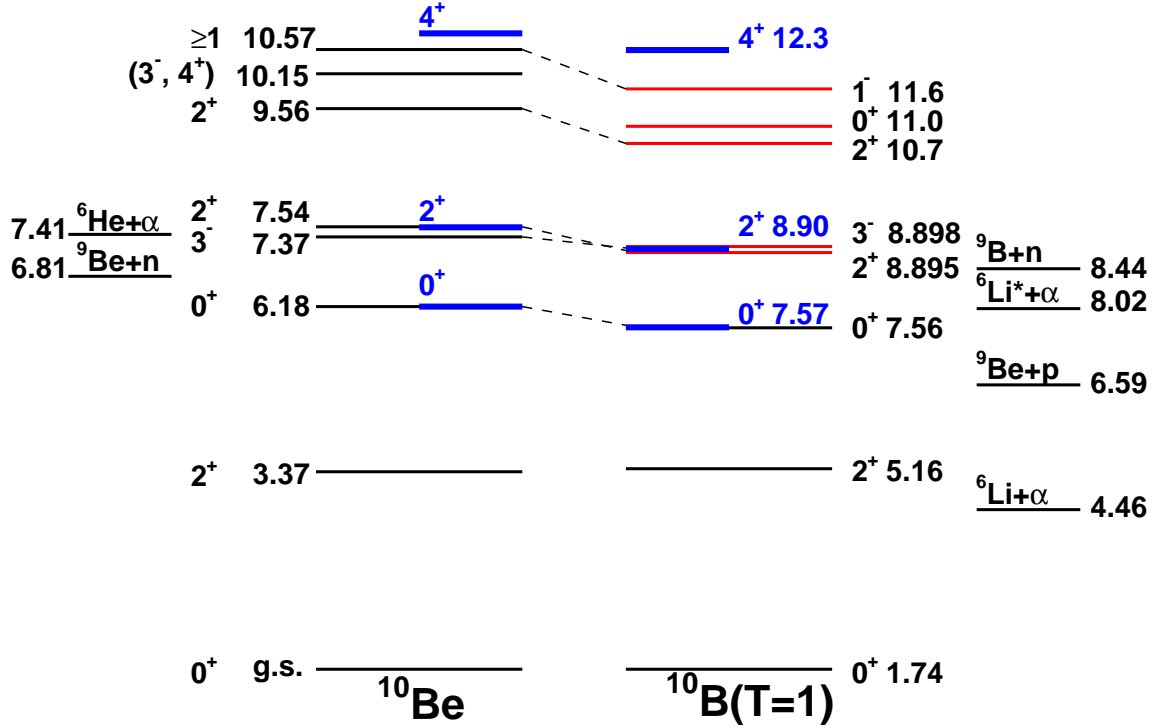}%
 \caption{\label{fig:levels} (Color online) Energy level diagram of select $^{10}$Be states and T=1 states of $^{10}$B. The states observed in this experiment are shown in red. Suggested isobaric analogs are connected by dashed lines. The energy levels generated by the $\alpha$+core potential model are shown in bold blue (see text for details).}
 \end{figure*}

One of the most interesting questions about the nuclear structure of $^{10}$Be and $^{10}$B is how the well known $\alpha$-cluster structure of $^8$Be manifests itself in the presence of two extra nucleons. As we discussed in the introduction, several theoretical approaches predict a developed $\alpha$:2n:$\alpha$ structure for the 0$^+$/2$^+$/4$^+$ band with 0$^+$ bandhead at 6.179 MeV in $^{10}$Be. The same should be true for the corresponding analog states in $^{10}$B. The most direct experimental observable for the degree of clustering is the partial $\alpha$ width. The excitation energies of the 0$^+$ states in both $^{10}$Be and $^{10}$B are below the corresponding $\alpha$ decay thresholds. The partial $\alpha$ widths for the 2$^+$ states were unknown in both nuclei (we consider the only available estimate for the partial $\alpha$ width of the 2$^+$ state at 7.54 MeV in $^{10}$Be \cite{Liendo2002}  unrealistic because it exceeds the $\alpha$ single particle limit by a factor of 30).
The 18$\pm$2.0$\pm$2.3(sys) keV partial $\alpha$ width for the 2$^+$ state at 8.89 MeV in $^{10}$B determined in this experiment corresponds to the $\alpha$ single particle limit. In fact, this width can be well reproduced in the framework of a simple $^{6}$Li(T=1)+$\alpha$ potential model. The Woods-Saxon potential with depth of -119 MeV, radius and charge radius of 2.58 fm and 2.27 fm respectively, and diffuseness of 0.677 fm generates the 0$^+$ and the 2$^+$ $\alpha$-cluster states in excellent agreement with the spectra of $^{10}$B and $^{10}$Be (see Fig. \ref{fig:levels}) and predicts the 15 keV width for the purely $\alpha$-cluster 2$^+$ state at 8.894 MeV in $^{10}$B (0.870 MeV above the $^6$Li(0$^+$,3.56 MeV)+$\alpha$ decay threshold). 

The partial $\alpha$ width and $\alpha$ decay branching ratio of the 2$^+$ at 7.54 MeV in $^{10}$Be can now be evaluated using isospin invariance. Taking into account the difference in penetrability factors between $^{10}$B and $^{10}$Be we estimate the branching ratio for $\alpha$ decay of the 2$^+$ state in $^{10}$Be to be $\Gamma_{\alpha}/\Gamma = 1.3\pm0.3\times10^{-4}$. This value (that already corresponds to the $\alpha$ single particle limit) is a factor of 30 smaller than the $\Gamma_{\alpha}/\Gamma=(3.5\pm1.2)\times10^{-3}$ reported for this state in \cite{Liendo2002}.

The very large partial $\alpha$ width of the 2$^+$ state at 8.89 MeV measured in this work leaves no doubt about its $\alpha$-$^6$Li(T=1) molecular type nature. This confirms the assertion made in several theoretical and experimental works \cite{Dote1997,Itagaki2000,Freer2006} that this state (or its analog at E$_x$=7.542 MeV in $^{10}$Be) is a member of a highly clustered rotational band built on the 0$^+$ state at E$_x$=7.56 MeV in $^{10}$B (6.179 MeV in $^{10}$Be). The defining feature of this band is its high moment of inertia which is indicative of the large separation between the two $\alpha$ cores \cite{Dote1997}. Assuming that the 10.15 MeV state observed in \cite{Freer2006} is the 4$^+$ member of the analogous band in $^{10}$Be it can be expected that the excitation energy of the corresponding 4$^+$ in $^{10}$B is $\sim$11.5 MeV. Indeed, if this state has a large dimensionless reduced $\alpha$ width as suggested in \cite{Freer2006} for the presumably analogous 10.15 MeV state in $^{10}$Be then there is a good chance to see it in the $^1$H($^9$Be,$\alpha$)$^6$Li$^*$(T=1) reaction. The 2$^+$ resonance at 8.89 MeV that is considered to be a member of the same rotational band as the aforementioned 4$^+$ is the dominant feature in the measured excitation function (Fig. \ref{fig:55deg}). However, the 4$^+$ state has not been observed in this work. Besides a trivial reason for not seeing this state because it does not exist or is at higher excitation energy, we can offer another explanation. If this state corresponds to pure $\alpha$+$^6$Li(T=1) molecular configuration and the admixture of the $^9$Be(g.s.)+p configuration is too small then the cross section for the (p,$\alpha$) reaction may be too small making this resonance ``invisible'' on top of the ``background'' of other T=1 states in $^{10}$B. This may also explain the results of \cite{Uroic2009} where excited states of $^{10}$B were populated in the $^{11}$B($^3$He,$\alpha$)$^{10}$B reaction and the 11.5 MeV state was only observed in the $^6$Li+$\alpha$ decay channel, and not in the $^9$Be(g.s.)+p channel. We estimate that the proton dimensionless reduced width for the $4^+$ state has to be less than $2\times 10^{-4}$ to make it unobservable in our measurements. This is 0.6\% of the proton dimensionless reduced width of the $2^+$ that corresponds to the same rotational band.  Its curious that the aforementioned potential model generates the 4$^+$ state in $^{10}$B at 12.3 MeV (see Fig. \ref{fig:levels}). The resonance structure at 11.6 MeV excitation in $^{10}$B is evident, but it corresponds to the 1$^-$ state with weak $\alpha$-cluster component. It is possible that this state is the isobaric analog for the 10.57 MeV state with uncertain ($\ge1$) spin-parity assignment in $^{10}$Be \cite{Tilley2004}.

Another unexpected result of this work is that the 3$^-$ state at E$_{cm}$=2.31 MeV (E$_x$=8.9 MeV) appears to be a highly clustered state as well ($\theta^2_{\alpha}$=0.42$\pm0.04$). Because of this high clustering, the ratio $(\Gamma_p{\Gamma}_{\alpha})^{2^+} /(\Gamma_p{\Gamma}_{\alpha})^{3^-}$ is significantly smaller (30 times) than previously reported in \cite{Kiss1977}. The origin of this is not clear, and further investigation is needed. This state is an isobaric analog of the 3$^-$ at E$_x$=7.371 MeV in $^{10}$Be that has long been identified with the $K^{\pi}$=1$^-$ rotational band with 1$^-$ bandhead at E$_x$=5.96 MeV. This band presumably corresponds to the (0s)$^4$(0p)$^5$(1s0d) shell model configuration  \cite{Bouten1970}  and members of this band should not have large overlap with the $\alpha\times^6$He(g.s.) configuration. Similar conclusions can be drawn from the results of the Molecular Orbit model of $^{10}$Be described in \cite{Itagaki2000}. The 3$^-$ state at E$_x$=7.371 MeV was suggested to belong to a group of 7 states formed by the coupling effect between the two bands, $K^{\pi}$=1$^-$ and $K^{\pi}$=2$^-$. These bands are formed by the two valence neutrons, one in the $\pi$-orbit and another in the $\sigma$-orbit with respect to the $\alpha$-$\alpha$ two-center system. The $K^{\pi}$=1$^-$ and $K^{\pi}$=2$^-$ bands are produced by states that have opposite valence neutrons spin projection (S$_z$=0) and the same (S$_z$=1) respectively. While $^6$He forms naturally in this scheme when the two neutrons are associated with the same $\alpha$ cluster, it cannot be in the 0$^+$ g.s. because the minimum spin in this case is 1. Therefore, the overlap between the wave function of the 3$^-$ state at E$_x$=7.371 MeV and the $^6$He+$\alpha$ configuration is expected to be small. The RGM study of the $\alpha$+$^6$He system performed in \cite{Fujimura1999} predicts the $K^{\pi}$=0$^-$ rotational band with large $\alpha$+$^6$He(g.s.) dimensionless reduced widths ($\sim$0.25 if calculated using channel radius of 6 fm) at higher excitation energy. Specifically, the 3$^-$ member of the $K^{\pi}$=0$^-$ band is suggested at 3.7 MeV above the $\alpha$ decay threshold (10.5 MeV excitation energy in $^{10}$Be). It is difficult to reconcile the results of the aforementioned theoretical studies with the experimental result of this work. Independent confirmation of the $\alpha$+$^6$He(g.s.) molecular nature of the 3$^-$ state can be obtained in $^6$He($^6$Li,d) or $^6$He($^7$Li,t) experiments. The very large $^6$He($^6$Li,d)$^{10}$Be$^*$($\sim$7.5 MeV) cross section measured in \cite{Milin1999} is indicative that at least one state in the 2$^+$/3$^-$ doublet is a highly clustered state but resolution was not sufficient to determine which one. Our result indicates that both states are highly clustered (although the 2$^+$ is still the dominant one) and it is desirable to perform $^6$He($^6$Li,d) or $^6$He($^7$Li,t) experiments with resolution better than 100 keV to confirm this.

The finding of a new broad 0$^+$ state at $\sim$11 MeV was surprising because there were no predictions for such a state. Its large partial $\alpha$ width is the direct evidence for the extreme $\alpha$ cluster structure of this state. Similar purely $\alpha$-cluster 0$^+$ states in the vicinity of 10 MeV excitation have been identified in several other light nuclei: in $^{12}$C at 11 MeV \cite{Fynbo2005}, in $^{18}$O at 9.8 MeV \cite{Johnson2009} and in $^{20}$Ne at 8.6 MeV \cite{McDermott1960,Shen1994}. Due to the pure $\alpha$+core nature of these states and large effective radius of the wave function that corresponds to these configurations (large separation of $\alpha$ from the core) these states are called ``$\alpha$-halo'' states in \cite{Funaki05}. Before this work these broad structures have been observed only in resonance scattering of $\alpha$ particles \cite{McDermott1960,Shen1994,Johnson2009} and in $\beta$-delayed $\alpha$-decay \cite{Fynbo2005}. The important result of this work is that the low lying very broad 0$^+$ $\alpha$-cluster level has evident nucleon width providing for population in (p,$\alpha$) resonance reaction.  We expect that the cross sections of a number of the reactions of the astrophysical importance can be affected by possible presence of levels with these properties. Obviously, this is intriguing and prompts further investigation. As we already pointed out  (sec. \ref{sec:0+}), we consider the broad T=1 0$^+$ state in $^{10}$B tentative. If the broad structure observed in this experiment is indeed the 0$^+$ state (and is not an artifact of the direct (p,$\alpha$) reaction and/or other broad low spin states, as discussed in the sec. \ref{sec:0+}) then it can also be observed in $^6$He+$\alpha$ elastic scattering. Excitation energy of this 0$^+$ state in $^{10}$Be should be $\sim$10 MeV, that is $\sim$3 MeV above the $\alpha$-decay threshold. Therefore, it is highly desirable to measure the excitation function of $^6$He+$\alpha$ resonance elastic scattering in a wide energy and angular range.

\section{Summary}

T=1 states in $^{10}$B have been studied by measuring the excitation function of the $^1$H($^9$Be,$\alpha$)$^6$Li*(0$^+$,3.56 MeV) reaction. Five T=1 states were used to make an R-matrix fit of the data. These states are shown in the energy level diagram in red in Fig. \ref{fig:levels} as well as possible analogs in $^{10}$Be connected by dashed lines. It was found that the 2$^+$ state at 8.9 MeV (isobaric analog of the 7.542 MeV state in $^{10}$Be) has a dimensionless reduced $\alpha$ width of 1.1$\pm0.2$. The high degree of clustering for this state confirms the predictions made in several theoretical studies \cite{Dote1997,Itagaki2000} that this state is the member of the highly deformed $\alpha$:np:$\alpha$ ($\alpha$:2n:$\alpha$ in $^{10}$Be)  molecular type rotational band.

The $^{10}$B isobaric analog of the $^{10}$Be 10.15 MeV state has not been found in this experiment, and no input on the 3$^-$/4$^+$ spin-parity assignment ambiguity for the state can be made. Absence of this state in the excitation function of $^1$H($^9$Be,$\alpha$)$^6$Li*(T=1, 0$^+$, 3.56 MeV) reaction sets an upper limit on the proton dimensionless reduced width for this state at $2\times 10^{-4}$ with 90\% confidence level (if excitation energy of this state in $^{10}$B is $\sim$11.5 MeV as expected for the isobaric analog of the $^{10}$Be 10.15 MeV state). The T=1 1$^-$ state at 11.6 MeV in $^{10}$B is observed. This state is probably an isobaric analog of the 10.57 MeV state in $^{10}$Be. The state at 10.84 MeV is identified as a T=1 isobaric analog of the 2$^+$ at 9.56 MeV in $^{10}$Be.

The evidence for a very broad 0$^+$ resonance at 11 MeV that corresponds to a pure $\alpha$+$^6$Li configuration is observed. Similar purely $\alpha$-cluster 0$^+$ states in the vicinity of 10 MeV excitation energy are found in $^{12}$C, $^{18}$O, $^{20}$Ne \cite{Fynbo2005,Johnson2009,McDermott1960}. Unfortunately, our results related to identification of the broad 0$^+$ state should be considered tentative because other explanations for the observed enhancement in the (p,$\alpha$) cross section may be possible (see discussion in section \ref{sec:0+}). Therefore, it is of particular interest to observe this structure in $^{10}$Be by measuring the excitation function of $^6$He+$\alpha$ elastic scattering.

The 3$^-$ state at 8.9 MeV (isobaric analog of the 7.371 MeV state in $^{10}$Be) was found to have a high degree of $\alpha$+$^6$Li(0$^+$) clustering ($\theta_{\alpha}=0.42\pm0.04$). This is surprising because it appears to contradict the conclusions of \cite{Bouten1970,Itagaki2000} that this state is the member of the $K^{\pi}$=1$^-$ rotational band. A more detailed study, both experimentally and theoretically, is needed to resolve this discrepancy.

\begin{acknowledgments}
The authors are grateful to Professor John Hardy for valuable comments and discussions. One author (AK) thanks Melina Avila for assistance with potential model calculations. The authors are also grateful to the staff at LNS-INFN for the excellent working condition and hospitality that was provided. This work was supported in part by the National Science Foundation under Grant Nos. PHY-456463 and PHY-1064819 and by the US Department of Energy Grant Nos. DE-FG02-93ER40773 and DE-FG52-06NA26207. 
\end{acknowledgments}

\bibliography{10B_November2011_preprint.bib}  
\end{document}